\documentclass[12pt]{revtex4}
\usepackage{times}
\usepackage{hyperref}
\usepackage[utf8]{inputenc}
\usepackage{amssymb,amsmath,amsthm, bbm, mathrsfs, xfrac, faktor,slashed}
\usepackage[T1]{fontenc}
\usepackage{graphicx}
\usepackage{color}
\usepackage{braket}
\usepackage{framed}
\usepackage[framemethod=tikz]{mdframed}
\colorlet{shadecolor}{gray!20}
\definecolor{refkey}{rgb}{0,0,1}
\definecolor{labelkey}{rgb}{1,0,0}

\theoremstyle{definition}

\numberwithin{equation}{section}

\newcommand{\C}{\mathbb{C}}

\newcommand{\arxiv}{arXive:}
\renewcommand{\ket}[1]{ #1\rangle}
\renewcommand{\bra}[1]{\langle #1}
\newcommand{\be}{\begin{equation}}
\newcommand{\ee}{\end{equation}}

\begin{document} 
\begin{flushright}
KCL-PH-TH/2019-30
\end{flushright}

\title{Fermionic spectral action and the origin of nonzero neutrino masses} 
\author{Mairi Sakellariadou}
\affiliation{Theoretical Particle Physics and Cosmology Group, Department of Physics, King’s College	London, University of London, Strand, London, WC2R 2LS, U.K.}
\email{mairi.sakellariadou@kcl.ac.uk}
\author{Andrzej Sitarz}
\affiliation{Institute of Physics, Jagiellonian University,
	prof.\ Stanis\l awa \L ojasiewicza 11, 30-348 Krak\'ow, Poland, \\ 
Institute of Mathematics of the Polish Academy of Sciences, \'Sniadeckich 8, 00-656 Warszawa, Poland.}
\email{andrzej.sitarz@uj.edu.pl}   
\thanks{Partially supported by Polish National Science Center (NCN) grant 2016/21/B/ST1/02438}
\pacs{02.90.+p, 12.90.+b, 14.60.Pq}
\begin{abstract} 
	We propose that the fermionic part of the action in the framework of the noncommutative description of the Standard Model is spectral, in an analogous way to the bosonic part of the action that is customary considered as being spectral. We then discuss the terms that appear in the asymptotic expansion of the fermionic spectral action.
\end{abstract} 
\maketitle 
\section{Introduction}

Current experimental data have confirmed neutrino oscillations, implying that at least two of the  neutrinos have small albeit nonzero masses \cite{PDG}.  Several models have been consequently proposed to explain the origin of such nonzero neutrino masses, considering neutrinos as being either Dirac fermions or Majorana particles \cite{Bil}. 

In the geometric interpretation of the Standard Model of particle physics, proposed within the framework of noncommutative geometry, neutrinos were originally 
massless Majorana particles \cite{Co95}. Yet the experimental confirmation that, beyond any doubt, neutrinos of different flavor oscillate, has enforced the introduction of the neutrino masses into the model. Most of the Standard Model phenomenology obtained through the noncommutative spectral geometry approach is based on considering either Dirac or Majorana massive neutrinos and employing the see-saw mechanism 
\cite{CCM}.

In this note we propose a consistent treatment of both the fermionic
and the  bosonic parts of the action \cite{Co97}, an approach which implies
nontrivial corrections leading to nonminimal fermion couplings. 
The idea behind the fermionic spectral action we propose, follows the customary
bosonic spectral action approach. For the bosonic spectral action, one assumes the existence of some energy cut-off
$\Lambda$, and thus obtains an effective action that depends on the 
spectrum of the Dirac operator truncated at $\Lambda$. One then
computes the asymptotic expansion of the leading term in 
$\Lambda$. We propose a similar approach for the fermionic action, 
usually expressed  as the expectation value of the 
Dirac operator for a fermionic field $\Psi$, which depends 
only on the truncated Dirac operator (namely one considers only the terms 
with eigenvalues smaller than the cutoff $\Lambda$).

The purpose of this paper is to investigate the qualitative consequences of
the spectral fermionic action we propose, leaving the details 
of the model (including the Lorentzian formulation
\cite{Bar}, fermion doubling \cite{LMM,AKL}, various forms of the fermionic 
action for different KO-dimensions etc) for a future study. Though usually,
it is argued that the bosonic spectral action offers a window into very high
energies \cite{Vas16} and implies a natural framework for an early universe cosmology \cite{ms},
we believe that correction terms to the fermionic spectral
action can be even currently observed. A previously proposed  
toy model \cite{Si09} suggested that the fermionic spectral action 
might be responsible for additional mass terms; in the following we discuss in detail 
the possible nonminimal interaction terms.

\section{The fermionic action}

The fermionic action for the spectral triple, which gives the dynamics and
interactions between fermions and the bosonic field, is usually formulated as 
the expectation value of the Dirac operator $D_A $ that takes the inner fluctuations into account, in the state given by a fermionic
field $\Psi$:
\begin{equation}
S_f = \bra{\Psi} \,|\, D_A \ket{\Psi}. 
\end{equation}
For the almost commutative geometries, with a specific KO dimension 2 (mod 8),
which is used in the description of the Standard Model,  one could consider another version 
of the action (on the total Hilbert space), in terms of the antisymmetric bilinear form:
\begin{equation}
S_{J,f} = \bra{J\Psi} \,|\,  D_A \ket{\Psi}, 
\end{equation}
where $J $ the real structure on the spectral triple defined by the choice of an algebra, a Hilbert space and a self-adjoint Dirac operator.
This form may be restricted to the subspace of right-handed fermions, reducing the
unnecessary doubling of the space of fermions \cite{LMM}.

In the noncommutative geometry approach there exists, however, a remarkable and quite unnatural difference
between the two parts of the action. While the bosonic part depends only on
the reduced spectrum of the Dirac operator, the fermionic one considers explicitly 
the full spectrum. This has been first observed in \cite{ACh98}, where it was postulated that also the fermionic action should have a similar form, yet the problem was not  discussed until the analysis of \cite{Si09}.

\section{The spectral fermionic action}

We propose to write the fermionic action functional, using similar cut-off regularization as is customary done for the bosonic 
case. Let us note that a zeta-function regularization for the bosonic spectral action has been proposed in 
\cite{Kurkov:2014twa} in order to address the issues of renormalizability and spectral dimensions.
Certainly, using the spectral action generalization leads to issues with the interpretation of the Euclidean formulation of spinors and the reduction of the model due to the so-called fermion doubling (or, more correctly, double doubling).

We shall work solely in the Euclidean setup, leaving aside potential problems and looking only for
terms that could lead to new physics. Our approach follows the one used for the bosonic spectral action, where the leading terms
introduce one by one the geometry, the interactions, and the coupling between them. The fermionic term, by 
definition gives only the minimal coupling between fermions, gauge fields and the Higgs. In what follows we will investigate
the type and form of nonminimal couplings, motivated by the geometric structure of the interactions
as described by noncommutative geometry.

Let us propose (for the simplest Euclidean model) the cutoff fermionic action
\begin{equation}
\label{fcsa}
S_{g,\Lambda} = \bra{\Psi} \,|\,  g_\Lambda\!\left( D_A \right) \ket{\Psi}, 
\end{equation}
for a suitable function $g$ and taking $g_\Lambda(x) = g(\frac{x}{\Lambda})$. Observe, that we take a function of $D_A$ and not  $D_A^2$ on purpose, as the spectrum of $D_A$ is not necessarily invariant with respect to change $D_A \to - D_A$.
Since any function can be splitted into the sum of an even and an odd function, 
and an even function can be taken as a function of $D_A^2$, and using
\be
\label{fheq}
g(x) = f(x^2) + x h(x^2),
\ee
one realizes that Eq.~(\ref{fcsa}) includes two choices:
\begin{equation}
 S_{e,\Lambda} =  \bra{\Psi} | \,f_\Lambda\!\left( D_A^2  \right) \ket{\Psi}, \qquad 
S_{o,\Lambda} = \frac{1}{\Lambda} \bra{\Psi} | \, D_A h_\Lambda\!\left( D_A^2 \right)  \ket{\Psi},
\end{equation}
where in principle, $f,h$ are two arbitrary functions of the cut-off type. Using the
fact that $D_A$ commutes with $h_\Lambda\!\left( D_A^2 \right)$  we can rewrite the latter expression as:
\begin{equation}
S_{o,\Lambda} =  \frac{1}{\Lambda} \bra{\Psi} | \, h_\Lambda\!\left( D_A^2 \right)  \ket{D_A \Psi},
 \end{equation}
which would enable us to use the tools of heat trace expansion \cite{Gi, Va03} for commutative and almost commutative geometries.
 
\subsection{Commutative geometries}

Let $M$ be a Riemannian spin manifold and $L^2(S)$ its spinor bundle and
$\Psi$ a spinor field. Define by $P_\Psi$  an endomorphism of the 
bundle of spinors that locally, at each point $x \in M$, projects on the spinor 
$\Psi(x)$, $ \Phi(x) \mapsto \Psi(x) \bra{\Psi(x)} | \ket{\Phi(x)}$, or equivalently, using physics notation $P_\Psi =   |\ket{\Psi(x)} \bra{\Psi(x)}|$. 

Let $D$ be a Dirac operator that lifts the torsion-free Levi-Civita connection to 
the spinor bundle. Although one can consider Dirac operators that arise from
connections with torsions, we concentrate on the usually assumed case
of vanishing torsion. 
We shall interpret the above fermionic action terms as arising from the  
heat kernel expansion of the type:
\begin{equation}
S_{e,\Lambda} = \hbox{Tr} \left( P_\Psi  g_\Lambda \left( D^2 \right) \right), 
\end{equation}
noting that $P_\Psi$ is a local operator (endomorphism of the bundle on which 
$D$ acts). Following the same approach as for the bosonic spectral action,  
we use the heat trace expansion:
\be
\hbox{Tr} \left(T e^{-t D^2} \right) = \sum_{n=0} t^{\frac{1}{2}(n-4)} \int_M a_n(x,T), 
\ee
where
\be
a_0(T,x) = (4\pi)^{-2} \hbox{tr}\, T, \qquad 
a_1(T,x) = (4\pi)^{-2} \hbox{tr} \left( T \left( -\frac{R}{6} + E \right) \right).
\ee
For the fermionic spectral action, the coefficients of this expansion are
the moments of the functions $f$ and $h$ \cite{Va03} (see Eq.~(\ref{fheq})). 
Here $\hbox{tr}$ denotes the local trace operation in the endomorphisms of
the spinor bundle taken at point $x$. Note that $\hbox{tr} P_\Psi = \bra{\Psi(x)}\ket{\Psi(x)}$ and therefore, for a $4$-dimensional manifold we shall 
have the following leading terms, which arise from the fact that for the Dirac 
operator $E = R/4$, with $R$ the scalar Ricci curvature: 
\begin{eqnarray}
\begin{aligned}
& \Lambda^4 \int_M \sqrt{g}\, (\hbox{tr} P_\Psi ) = \Lambda^4 \int_M \sqrt{g} \; 
\bra{\Psi(x)}  | \ket{\Psi(x)}, \\
& \Lambda^2 \int_M \sqrt{g}\, \hbox{tr} \left( P_\Psi \frac{R}{12} \right)
= \Lambda^2 \int_M \sqrt{g} \,  \frac{R}{12} \; \bra{\Psi(x)} | \ket{\Psi(x)}.
\end{aligned}
\end{eqnarray}
The leading term (in $\Lambda^4$), which resembles the cosmological constant term, corresponds to the bare fermion mass term whereas the second term (in $\Lambda^2$) comes from the even part of spectral the action and describes the non-minimal coupling of fermions to gravity through the scalar curvature \cite{CII,BD}.

As for the odd part of the fermionic spectral action, we propose to write it 
using similar arguments as:
\be
S_{o,\Lambda} = \frac{1}{\Lambda} \hbox{Tr} \left( P_{\Psi, D \Psi} \,  h_\Lambda
\left( D^2 \right) \right), 
\ee
where the local operator $P_{\Psi, D\Psi}$ is and endomorphism of the 
spinor bundle of the form:
\be
\Phi(x)\mapsto D\Psi(x) \bra{\Psi(x)} | \ket{\Phi(x)}. 
\ee
In a similar way as for the even case, observe that $\hbox{tr} \, P_{\Psi, D\Psi}
= \bra{\Psi(x)}| \, D \ket{\Psi(x)}$
with $(,)$ the local scalar product on the sections of the spinor bundle, with 
values in $C^\infty(M)$.  

The odd component of the fermionic spectral action introduces at leading order (in $\Lambda^3$) the fermion dynamics: 
\be
\Lambda^3 \int_M \sqrt{g}\,  \bra{\Psi(x)} | D \ket{\Phi(x)}, 
\ee
while the next order terms would contribute (in the case of pure gravity) further coupling
to scalar curvature $R$: 
\be
\Lambda \int_M \sqrt{g}\,  \frac{R}{12} \,  \bra{\Psi(x)} | D \ket{\Phi(x)}. 
\ee

\subsection{The Einstein-Yang-Mills system}

The above discussed case extends to the situation of a simple noncommutative modification of geometry \cite{Co97} where one considers the algebra of 
$M_n(\mathbb{C})$ valued functions on the spin manifold $M$ and uses the 
family of Dirac operators constructed by gauge fluctuation of the Dirac operator 
$D$. Such family, which is obtained through the so-called internal fluctuations 
of the metric, is of the form:
\be
D_A = D \otimes \hbox{id} + A, 
\ee
where $A$ is a gauge potential ($A = \sum_i a_i [D,b_i]$) for $a_i,b_i \in C^\infty(M) \otimes M_n(\mathbb{C})$. This includes, in particular, the case of Dirac operators
twisted by a connection on a vector bundle. 

If the spectral triple is real and satisfies the order-one condition \cite{CoMa} one 
should modify the above family by correcting $A$ through (with a sign 
dtermined by $JD= \pm DJ$) $ J A J^{-1}$. Since the square of the 
Dirac operator contains the gauge curvature $F$, the formulae for the heat 
trace expansion are modified, $E = (R/12) + F$, and consequently the first three 
leading fermionic action terms are modified as follows:
\begin{eqnarray}
\begin{aligned}
& \Lambda^4 \int_M \sqrt{g}\, (\hbox{tr} P_\Psi ) = \Lambda^4 N \int_M \sqrt{g} 
 \bra{\Psi(x)} | \ket{\Psi(x)}, \\
& \Lambda^3 \int_M \sqrt{g}\, \bra{\Psi(x)} | D_A \ket{\Psi(x)}, \\
& \Lambda^2 \int_M \sqrt{g}\, \bra{\Psi(x)} | \left( \frac{R}{12} + F \right)  \ket{\Psi(x)}, 
\end{aligned}
\end{eqnarray}
The only difference from the previous case is -- apart from the multiplicity of fermionic fields that comes from the representation of the algebra $M_N(\mathbb{C})$ -- the 
minimal coupling of fermions with gauge fields that appears in the second 
term (in $\Lambda^3$) as well as the Pauli interaction Lagrangian that appears in the third term (in $\Lambda^2$), that locally looks like
\be
\int_M \sqrt{g} \,  \bra{\Psi(x)} | F \ket{\Psi(x)}= \int_M \sqrt{g} \; \Psi(x)^\dagger \left( \gamma^\mu \gamma^\nu F_{\mu\nu}(x) \right) \Psi(x),
\ee
and which can be nontrivial even in the case of electrodynamics, which in the case when $J$ is present corresponds to the algebra $\C \oplus \C$.

\subsection{The almost commutative geometries}

Let us now consider the simplest noncmmutative geometry, made from the product of a smooth four-dimensional manifold $M$ (with a fixed spin structure), by a discrete finite-dimensional noncommutative internal space $F$ , defined in the language of finite spectral triples. Since effectively the  Dirac operator of this product geometry is of the same type as for an Einstein-Yang-Mills system,  
one does not expect any significant qualitative differences from the corresponding 
studies of commutative geometries.

Let us recall the basic notions. We take as the underlying algebra of the model
$\mathcal{A} = C^\infty(M) \otimes \mathcal{A}_F$,  which is represented 
on the Hilbert space $L^2(S) \otimes \mathcal{H}_F$, and the Dirac operator is 
$\mathcal{D} = D_M \otimes \hbox{id} + \gamma_5 \otimes D_F$, where again $D$ denotes 
the standard Dirac operator on the spin manifold $M$ and $D_F$ is the Dirac 
operator of the finite spectral triple
$(\mathcal{A}, \mathcal{H}, D_F)$. We refer for the details of construction of 
product geometries and related issues to \cite{Co95}.

The family of Dirac operators $\mathcal{D}_{\mathbb{A}}$ arises similarly as discussed in the previous section, as fluctuations of the Dirac operator. They include both the classical gauge fields, with the unitary group of inner automorphisms of the algebra ${\mathcal{A}_F}$ as well
as the gauge fields related to discrete geometry, which are interpreted as the
Higgs field. More precisely,
\be
\mathcal{D}_{\mathbb{A}} = D \otimes \hbox{id} + \mathbb{A} 
+ (\gamma^5 \otimes 1)D_F(H), 
\ee
where $\mathbb{A}$ are the  inner fluctuations of 
the Dirac operator $D$ (containing, if we assume the reality of the spectral 
triple, also the real part of fluctuation) and $D_F(H)$ are  inner fluctuations of the product geometry with respect to the discrete Dirac operator $D_F$. 
Note that both $\mathbb{A}$ and $D_F(H)$ are, from the technical point of 
view, just matrix-valued functions on the manifold $M$, which are represented
on $L^2(S) \otimes \mathcal{H}_F$. 
 
To obtain the leading terms in the spectral action we use the heat trace
asymptotic expansion for the square of the Dirac operator, 
\be
{\mathcal{D}}_{\mathbb{A}}^2=\nabla^\star\nabla - E,
\ee
where $\nabla$ is a connection on the spinor bundle tensored with $\mathcal{H}_F$
and $E$ is the endomorphism of the latter bundle. 

In local coordinates over the manifold, with $\gamma^\mu$ being the usual gamma
matrices, we have:
\be
E = - (D_F(\Phi))^2 -\sum_{\mu<\nu}\gamma^\mu\gamma^\nu \mathbb{F}_{\mu\nu}
+ i \gamma_5\gamma^\mu \mathbb{M}(D_\mu(H)),
\ee
where $\mathbb{F}_{\mu\nu}$ is the curvature tensor of the gauge connections,
$(D_F(\Phi))^2$ is the potential term for the fields $H$ and the last term 
$\mathbb{M}(D_\mu(H)$ is the endomorphism of the bundle that depends on 
the covariant derivative of fields $H$.

We shall analyze these terms in the next section, in the particular case of the 
almost commutative geometry underlying the Standard Model.

\section{The application to the Standard Model} \label{secIV}

Let us briefly recall the basics of the Standard Model description within the framework of almost commutative geometry. To obtain the Standard Model the minimal choice of the algebra in the spectral triple defining the discrete internal space $F$ is $\mathcal{A}_F = \mathbb{C}
\oplus \mathbb{H} \oplus M_3(\mathbb{C})$. This algebra is represented on a 
16-dimensional Hilbert space that includes all fermions (assuming Dirac neutrinos)
or 15-dimensional if we work with Majorana neutrinos only. For the details of the
action in a convenient basis see \cite{CoMa} or \cite{FB, DDS18} for a most recent
formulation and principles of constructing the Dirac operator both for the quark
and leptonic sector.

The discrete Dirac operator written in the basis of fermions, taken in the order 
(for leptons) $\nu_R,e_R,\nu_L,e_L$ (not that as a rule each fermion is denoted
cumulatively for $N$ generations) is
\be
D_F = \left( 
\begin{array}{cccc}
 0 & 0 & \Upsilon_\nu^* &  0\\
 0 & 0 &  0& \Upsilon_e^* \\
\Upsilon_\nu & 0 & 0& 0 \\
0& \Upsilon_e  & 0& 0 \\
\end{array}
\right),
\ee
where $\Upsilon_\nu,\Upsilon_e$ are $N\!\times\!N$ mass and mixing
matrices. The fluctuated discrete Dirac operator $D_{F,H}$ is:
\be
D_{F,H} = \left( 
\begin{array}{cccc}
0 & 0 & \Upsilon_\nu^* H^0 & \Upsilon_\nu^* H^-  \\
0 & 0 &   -\Upsilon_e^*  \overline{H^-}  & \Upsilon_e^* \overline{H^0} \\
\Upsilon_\nu  \overline{H^0} &  -\Upsilon_e H^-& 0 & 0 \\
\Upsilon_\nu  \overline{H^-} & \Upsilon_e  H^0 & 0 &  0 \\
\end{array}
\right),
\ee
where $H = H^0 + H^-  j$ denotes a quaternionic field (Higgs doublet). The 
discrete part of the Dirac operator has such form also for the quark sector, 
if  we take quarks in the order $q_R^{u}, q_R^{d}, q_L^{u}, q_L^{d}$  and the mass and mixing matrices are,  respectively $\Upsilon_u$ and $\Upsilon_u$. 

\subsection{The fermionic spectral action for the SM}

As we have previously seen, the first two leading terms of the fermionic
spectral action are the bare mass term and the usual fermionic action.  
As the Standard Model is chiral, in the Lorentzian version the bare mass 
term is not possible as it is not gauge invariant. In the Euclidean version, 
however, it can appear in the model with the fermion doubling but will 
have to vanish if one requires that the action terms are restricted to the physical space of fermions (by removing the fermion doubling \cite{LMM}).

The next term yields the usual part of action, which includes the dynamical term for
fermions, minimal coupling to gauge fields and the coupling between the Higgs
field and fermions, which gives the mass terms in the broken symmetry phase.

The only possible corrections and new effects can be therefore visible in the
term, which is proportional to $\Lambda^2$.  Of course, we shall have 
there similar terms as in the Einstein-Yang-Mills system, that is the nonminimal
coupling of fermions to gravity (through the scalar curvature) and the Pauli-type
interaction terms (coupling to curvature of connections) \cite{RRS}.

However, we shall additionally have the  term of the fermionic spectral action that contains the square of the fluctuated discrete part of the Dirac operator, $D_F(H)^2$, that contains the Higgs field. We discuss now two interesting cases of Dirac and Majorana neutrinos, concentrating our analysis on the leptonic sector. Observe
that the same could be, of course, used for the spatial part of the Dirac manifold
leading to higher-derivative terms in the fermionic action that have been considered
in some models \cite{EGO00, Vil02}

\subsection{Dirac neutrinos}

Within the Standard Model, the square of the finite Dirac operator gives the corrections to the fermion masses. Using the notation we have introduced previously, we compute $D_F(\Phi)^2$ restricted to the lepton sector: 
\be
(D_{F,H_{l}})^2 = \left( 
\begin{array}{cccc}
\Upsilon_\nu^* \Upsilon_\nu |H|^2 + \Upsilon_R^* \Upsilon_R^* & 0 & 0 & 0\\
0 & \Upsilon_e^* \Upsilon_e |H|^2 & 0 & 0 \\
0 & 0 & \Upsilon_\nu^* \Upsilon_\nu |H|^2 & 0 \\
0 & 0 & 0 &  \Upsilon_e^* \Upsilon_e |H|^2
\end{array}
\right),
\ee
where $\Upsilon_e, \Upsilon_\nu$ are the mass and mixing matrices, 
respectively.

The order of the corrections, when compared to the main mass term are
of the order $1/\Lambda$ and therefore will be negligible when 
compared to the Dirac mass terms computed in the vacuum expectation
value of the Higgs field: $H = H_v$ (that is $H^0 = H_v, H^- = 0$).

\subsection{Majorana neutrinos}
In the noncommutative description of the Standard Model that uses only 
left-handed neutrinos there is no room for the neutrino mass terms. The
natural interpretation of such model is in terms of Majorana neutrinos, as
after restricting it to the physical subspace (reducing the fermion 
doubling) the neutrino spinor fields are 
their own antiparticles.
Of course, there are known mechanisms to generate possible mass terms,
yet all of them involve quadratic coupling to the Higgs. In view of the 
analysis of the fermionic spectral action we discussed previously, we argue below that one
can obtain such terms from the next leading term in the heat kernel expansion.

Observe, that if there are no right-handed neutrinos, the fluctuations
of the discrete Dirac operators on the leptonic sector, in the basis
$(e_R, \nu_L, e_L)$, are:
\be
D_{F,H} = \left( 
\begin{array}{ccc}
0 &   -\Upsilon_e^*  \overline{H^-}  & \Upsilon_e^* \overline{H^0} \\
- \Upsilon_e H^-& 0 & 0 \\
\;\; \Upsilon_e  H^0 & 0 & 0 \\
\end{array}
\right),
\ee
Note that by taking the term with the square of the finite part of the Dirac operator 
would not give anything new. Indeed, computing $(D_{F,H})^2$ at the Higgs vacuum 
expectation value  we obtain:
$$ 
(D_{F, H_v})^2 = 
\left( 
\begin{array}{ccc}
\Upsilon_e^* \Upsilon_e |H_v|^2  &  0& 0 \\
 0 & 0 & 0 \\
 0 & 0 & \Upsilon_e^* \Upsilon_e |H_v|^2  \\
\end{array} \right),
$$
which similarly as in the Dirac masses case can only add small corrections
to the already nonvanishing mass of charged leptons.

To see that some extra terms are possible we need to generalize the form
of the spectral action to nonscalar functions. So far we have assumed that
the function $f_\Lambda$, which we have taken to define the even part
of the spectral action is scalar, that is for every operator $T$ that 
commutes with $D_A$ we assume that $f_\Lambda(D_A)$ commutes with $T$ as
well. However, this is not necessary and we may consider other functions
provided the full gauge invariance will be preserved. 

Leaving aside the question about the classification of such functions,
for the specific model, we observe the existence of a particular one.
Let $\tau$ be the operator mapping $(\nu_L, e_L)$ to $(e_L^c, -\nu_L^c)$
where $(\nu_L^c,e_L^c)$ denotes the respective antiparticles.  Taking as
$f_\Lambda^\tau(D_A) = \tau f_\Lambda(D) \tau$ we still obtain a selfadjoind
operator provided that $\tau$ is selfadjoint (or antiselfaidjoint, as is the case 
for chosen $\tau$).

The operator $\tau$ could be written using Pauli matrices as $i \sigma_2 \circ J$, 
where $J$ is the reality operator of the model (see \cite{Co95,CCM}) restricted 
to the leptonic  sector. The action of the quaternionic part of the algebra 
on the antiparticle sector 
in the noncommutative description of the Standard Model is  through $\bar{h}$
for $h \in {\mathbb{H}}$, where quaternions are represented as complex matrices
in $M_2(\mathbb{C})$ and $\bar{h}$ denotes complex conjugated matrix. 
Since for any any quaternion we have $\tau h = \bar{h} \tau$ then 
$D_{F,H} \tau (e_L^c, \nu_L^c)$ is invariant under the $SU(2)$ part of 
gauge transformations. 

Writing explicitly in the  $(e_R^c, \nu_L^c, e_L^c)$ basis the matrix 
elements of $D_{F,H}\tau$:
\be
D_{F,H} \tau = \left( 
\begin{array}{ccc}
0 &    - \Upsilon_e^*  \overline{H^0}  & - \Upsilon_e^* \overline{H^-} \\
 \;\;     \Upsilon_e H^0& 0 &  0 \\
- \Upsilon_e  H^- & 0 &  0 
\end{array}
\right).
\ee
Then the terms in the fermionic spectral action, that arise from 
$\hbox{Tr}(P_\Psi \, f_\Lambda^\tau({\mathcal{D}}^2)$ in the next-to-leading
order, could be explicitly rewritten as:
\be
\Lambda^2 (\Upsilon_e {\Upsilon_e}^*)
\left[  
\left( \overline{\nu_L^c}, \overline{e_L^c} \right) 
\left( 
\begin{array}{c}
 \;\;      H^0 \\
  H^- 
\end{array}
\right)
\right]
\left[  
\left(  \overline{H^0} ,  \overline{H^-}\right) 
\left( 
\begin{array}{c}
\nu_L^c \\
e_L^c
\end{array}
\right)
\right] \; + \; \hbox{h.c}. 
\ee
As we have observed before the entire expression is gauge invariant and can be
identified as a Weinberg term (sometimes called Weinberg operator) \cite{Wei}, which is used to describe effective mechanism of neutrino mass generation. As the operator
is, in fact, non-renormalizable, one often explains the physics behind the effective 
term as originating from yet unknown heavy intermediate particles.

After the Higgs field gets its vacuum expectation value, which in our choice of the
parametrization is $H^0=H_v, H^-=0$, a neutrino mass is generated,
depending on the scale $\Lambda$, Higgs vacuum expectation value $H_v$, 
masses of charged leptons and the coefficients of the cutoff function 
$f_\Lambda^\tau$. Recall, however, that the usual mass terms appear
at order $\Lambda^3$, compared to $\Lambda^2$ for the Weinberg term.  
Therefore if one assumes $\Lambda$ to be much larger than the Higgs vacuum 
value (in many models this is around the scale of GUT)  then the generated neutrino masses will necessarily be small, which agrees with the experiment.
 
\section{Conclusions and outlook beyond the Standard Model}

As we have shown, even the simplest model, which is based on the almost commutative geometry with the finite part described by a spectral triple, leads
to the appearance of correction terms that give some non-minimal interactions
between gravity and fermions, gauge fields and fermions as well as the
Higgs field and fermions. It is interesting that no extra particles are required to 
explain the appearance of neutrino masses.
Of course, we have demonstrated only that the spectral action for fermions in the Euclidean version could be constructed and computed leaving aside the problem whether any restriction could appear when considering the Lorentzian version,
in particular with respect to the fermion doubling problem \cite{LMM}. 

It is also interesting that the next order corrections lead to terms that have been considered in various models, including Pauli interaction terms and nonminimal coupling to gravity.  Connecting their origins to the same spectral action principle as in the case of neutrino masses could help to set possible  limits on their observational evidence or set constraints on the models from cosmological observations in a 
similar way it can be done for the bosonic action \cite{NS10, NOS10}.

The neutrino mass corrections are possible in both Dirac and Majorana neutrino 
models, and may lead to small neutrino masses. The correction terms are nonrenormalizable  (which is similar to higher order terms from the bosonic action), yet could be treated as an effective description. 

The model allows an extension of the assumed form of the cutoff function to include also a nonscalar part, which means that some internal permutations of the
eigenspaces that are within the range of the spectral projection $P_\Lambda$ are
allowed. This point certainly requires further studies, as it is necessary to understand the allowed freedom in the choice of the function. In particular, to introduce the
neutrino mixing matrix one needs to generalize the cutoff function further, by
adding a mixing to the function (that is modifying $\tau$ operator so that it is not
diagonal for the three families).

\subsection*{Acknowledgements}
The authors would like to thank the referee for remarks and comments. The research was initiated thank to support of the STSM exchange, COST action {\em Qspace}.
Publication is supported by the John Templeton Foundation Grant, Conceptual Problems in Unification Theories” (No. 60671). MS is supported in part by the Science and Technology Facility Council (STFC), UK under the research grant ST/P000258/1.


\begin{thebibliography}{1}
\itemsep=4pt

\bibitem{PDG}
C. Patrignani et al. (Particle Data Group), 
\textit{Neutrino Masses,Mixing, and Oscillations},
Chin. Phys. C, 40, 100001 (2016) and 2017 update,

\bibitem{Bil}
S. M. Bilenky,
\textit{Neutrino in Standard Model and beyond},
Phys. Part. Nuclei (2015) 46: 475.



\bibitem{Co95} 
A. Connes, \textit{Gravity coupled with matter and the foundation of noncommutative geometry}, Commun. Math. Phys. 183, (1995), 155--176

\bibitem{CCM}
A.H. Chamseddine, A. Connes and M. Marcolli, 
\textit{Gravity and the standard model with neutrino mixing}, 
Adv. Theor. Math. Phys. 11 (2007) 991


\bibitem{Co97} 
A. Chamseddine, A. Connes, \textit{The Spectral Action Principle}, Comm. Math. Phys. 186, 731-750 (1997).

\bibitem{Bar}
J.W. Barrett,
\textit{Lorentzian version of the noncommutative geometry of the standard model of particle physics},

\bibitem{LMM}
F. Lizzi, G. Mangano, G. Miele and G. Sparano, 
\textit{Fermion Hilbert space and fermion doubling in the noncommutative geometry approach to gauge theories}, 
Phys. Rev. D 55 (1997) 6357

\bibitem{AKL}
A.A.Andrianov, M.A.Kurkov, F.Lizzi,  
\textit{Spectral action, Weyl anomaly and the Higgs-dilaton potential},
J. High Energ. Phys. (2011) 2011: 1.

\bibitem{Vas16}
D.Vassilevich,
\textit{Can Spectral Action be a Window to Very High Energies?}, 
J. Phys.: Conf. Ser. 670 012050

\bibitem{ms}
  M.~Sakellariadou,
 \textit{Aspects of the Bosonic Spectral Action: Successes and Challenges},
  PoS CORFU {\bf 2015}, 115 (2016)
; M.~Sakellariadou, \textit{Noncommutative Spectral Geometry: A Short Review},
  J.\ Phys.\ Conf.\ Ser.\  {\bf 442}, 012015 (2013).
  
\bibitem{Si09}
A.~Sitarz, \textit{Spectral action and neutrino mass}, EPL, 86, 10007,  (2009) 

\bibitem{ACh98} A.~Chamseddine, 
\textit{Remarks on the spectral action principle}, Phys.Lett. B 436, 84--90, (1998)

\bibitem{Kurkov:2014twa} 
  M.~A.~Kurkov, F.~Lizzi, M.~Sakellariadou and A.~Watcharangkool,
 \textit{Spectral action with zeta function regularization},
  Phys.\ Rev.\ D {\bf 91}, no. 6, 065013 (2015)

\bibitem{Gi}
P.B. Gilkey: \textit{Invariance theory, the heat equation, and the Atiyah-Singer index theorem}, Publish or Perish Press, 1985.

\bibitem{Va03} D. V. Vassilevich, \textit{Heat kernel expansion: user's manual'},
Phys. Rept. {388} (2003), 279--360.

\bibitem{FB}
S.Farnsworth, L.Boyle,
\textit{Rethinking Connes' approach to the standard model of particle physics via non-commutative geometry},
New J. Phys. 17, 023021 (2015)

\bibitem{DDS18}
L.~Dabrowski, F.~D'Andrea and A.~Sitarz, 
\textit{The Standard Model in noncommutative geometry: fundamental fermions as internal forms}, Lett Math Phys (2018) 108: 1323 

\bibitem{CII}
O. Cata A. Ibarra S. Ingenhutt
\textit{Dark matter decays from non-minimal coupling to gravity},
Phys. Rev. Lett. 117, 021302 (2016)

\bibitem{BD}
David G. Boulware, S. Deser
\textit{Canonical analysis of the fermion sector in higher-derivative supergravity},
Physical Review D, 30 (1984). pp. 707-71


\bibitem{CoMa}
A. Connes, M. Marcolli: \textit{Noncommutative Geometry, Quantum Fields and Motives}, Colloquium Publications, Vol. 55, Providence. RI: Amer. Math. Soc., 2008.



\bibitem{RRS}
R.R. Sastry,
\textit{Quantum Electrodynamics with the Pauli Term}
\arxiv{hep-th/9903179}

\bibitem{EGO00}
E. Elizalde, S. P. Gavrilov, S. D. Odintsov, Yu. I. Shilnov,
\textit{Dynamical Symmetry Restoration for a Higher-Derivative Four-Fermion Model in an External Electromagnetic Field},
Brazilian Journal of Physics, vol. 30, no. 3, September, 2000,

\bibitem{Vil02} Eduardo J S Villasenor
\textit{Higher derivative fermionic field theories} 
J. Phys. A: Math. Gen. 35 (2002) 6169–6182

\bibitem{Wei} S. Weinberg, 
\textit{Baryon and Lepton Nonconserving Processes.} 
Phys. Rev. Lett., 43, (1979), 1566--1570 

\bibitem{NS10}
W. Nelson and M. Sakellariadou, 
\textit{Cosmology and the noncommutative approach to the standard model}, 
Phys. Rev. D 81 (2010) 085038

\bibitem{NOS10}
W. Nelson, J. Ochoa and M. Sakellariadou, 
\textit{Constraining the noncommutative spectral action via astrophysical observations}, 
Phys. Rev. Lett. 105 (2010) 101602 

\end{thebibliography}
\end{document}